\RequirePackage{ifpdf}
\ifpdf 
\documentclass[pdftex]{sigma}
\else
\documentclass{sigma}
\fi

\begin{document}

\renewcommand{\PaperNumber}{003}

\FirstPageHeading

\ShortArticleName{Transient Phenomena in Quantum Bound States
Subjected to a Sudden Perturbation}

\ArticleName{Transient Phenomena in Quantum Bound States\\
Subjected to a Sudden Perturbation}

\Author{Marcos MOSHINSKY and Emerson SADURN\'I}

\AuthorNameForHeading{M. Moshinsky and E. Sadurn\'\i}

\Address{Instituto de F\'{\i}sica,
 Universidad Nacional Aut\'onoma de M\'exico,
  Apartado Postal  20-364,\\
 01000 M\'exico D.F., M\'exico}
\Email{\href{mailto:moshi@fisica.unam.mx}{moshi@fisica.unam.mx},
\href{mailto:sadurni@fisica.unam.mx}{sadurni@fisica.unam.mx}}

\ArticleDates{Received June 24, 2005, in final form August 08, 2005; Published online August 18, 2005}

\Abstract{Transient phenomena in quantum mechanics have been of interest to
one of the authors (MM) since long ago and, in this paper, we focus
on the problem of a potential~$V_-$ which for negative times gives
rise to bound states and is suddenly changed at $t=0$ to a~potential
$V_+$ which includes $V_-$ plus a perturbed term. An example will be
the deuteron (where the proton and neutron are assumed to interact
through an oscillator potential) submitted to a sudden electrostatic
field. The analysis for $t>0$ can be carried out with the help of
appropriate Feynmann propagators and we arrive at the result that
the separation between the nucleons has an amplitude that depends on
the intensity of the electrostatic field, but its period continues
to be related with the inverse of the frequency of the oscillator
proposed for the interaction. A general approximate procedure for
arbitrary problems of this type is also presented at the end.}

\Keywords{transient phenomena; propagators}

\Classification{81V35; 81Q05}


\section{Introduction}

The word ``Transient'' will mean for us here that the expectation
values of measurable observable or transition probabilities in
quantum mechanics will be time dependent, in contrast with the
``stationary'' case in which they are not. From the very origin of
quantum mechanics, in the ninety twenties, these transient phenomena
could be discussed through the time dependent versions of the
equations of motion in the Heisenberg or Schr\"odinger picture.

The most natural way of discussing the transient phenomena is
through the version of quantum mechanics that Feynman~\cite{1}
proposed in the ninety forties. For him the fundamental concept was a
propagator, also sometimes called Feynmans kernel, that relates an
initial state characterized by the wave function $\psi ({\boldsymbol x}', t')$
(where ${\boldsymbol x}'$ is a notation for all the coordinates of the system
and $t'$ its time) with the final state $\psi ({\boldsymbol x}, t)$ with
$t>t'$. This propagator was denoted by $K ({\boldsymbol x}, t; {\boldsymbol x}', t')$ and
Feynman gives a procedure to calculate it starting from a Lagrangian
formulation (in terms of generalized coordinates and velocities) of
the classical problem and summing all paths suggested by the action
principle between ${\boldsymbol x}'$ and ${\boldsymbol x}$. If the classical Lagrangian,
is not explicitly dependent on time, the propagator only depends on
the difference of the times i.e. $t-t'$ so, without loss of
generality, we could take $t'=0$ and write
\begin{gather*}
     K ({\boldsymbol x}, t; {\boldsymbol x}', 0) \equiv K ({\boldsymbol x},t; {\boldsymbol x}'), \qquad \psi
     ({\boldsymbol x}', 0) \equiv \psi ({\boldsymbol x}')
        \label{me01}
\end{gather*}
\pagebreak

\noindent
        so that we have
\begin{gather}
                \psi ({\boldsymbol x}, t) = \int K ({\boldsymbol x}, t, {\boldsymbol x}') \psi
                ({\boldsymbol x}') d{\boldsymbol x}'
        \label{me02}
        \end{gather}
which implies that the propagator $K ({\boldsymbol x}, t, {\boldsymbol x}')$ satisfies a
time dependent Schr\"odinger equation with the initial condition
                                    \begin{gather*}
        K ({\boldsymbol x}, 0; {\boldsymbol x}') = \delta ({\boldsymbol x}- {\boldsymbol x}').
        \label{me03}
        \end{gather*}

The use of the propagator concept has been enhanced by the
appearance a few years ago of a ``Handbook of Feynman Path
Integrals''~\cite{2}, and the power of the concept, in the case of a
one dimensional free particle, is illustrated in the Appendix  for a
compact derivation of the problem of ``Diffraction in time'', which
one of the authors (MM)~\cite{3} analyzed long ago by other methods.

The general type of problems that we wish to discuss here is that of
the potential $V_-({\boldsymbol x})$ which  for negative times $t<0$  is one
that admits bound states and that at $t=0$ is suddenly changed to a
potential $V_+({\boldsymbol x})$ that contains $V_-({\boldsymbol x})$ plus some type of
external interaction given by $V_+({\boldsymbol x}) - V_- ({\boldsymbol x})$. If our
vector ${\boldsymbol x}$ is given in cartesian coordinates the kinetic energy
is a sum of terms $(m_s \dot x^2_{qs})$ where $q=1,\dots, m$ is the
index for the dimension $m$ of our single particle vector space and
$s=1, \dots ,n$ the index for the number of particles. Thus our
classical Lagrangian is well defined and we can look through the
tables of reference~\cite{2} to see whether the propagator is
available. We thus proceed to analyze the following problem.

\section{The deuteron subjected to a sudden electrostatic field}

The first physical problem that came to our attention along the
lines of the last paragraph of the previous section, was the
hydrogen atom in a sudden constant electrostatic field.
Unfortunately the propagator for the Coulomb problem alone is
already quite complicated, and more so if a~linear electrostatic
potential is added.

We thus wanted to focus on problems in which the propagator can be
fully and simply determined, and that led us to Lagrangians that are
quadratic in the coordinate and velocities variables, of which there
is an extensive list of propagators in reference~\cite{2}.

Why are quadratic Lagrangian (which imply also quadratic
Hamiltonians in the coordinate and momentum variables) give
propagators of the Gaussian type~\cite{2}? One answer is given in a
paper by one of the authors (MM) and C.~Quesne~\cite{4}. The time
evolution associated with a~classical Lagrangian is given by a
canonical transformation which conserves a symplectic metric. For
quadratic Lagrangians or Hamiltonian this canonical transformation
is linear and it provides the dynamical group of the problem. To
translate it to quantum mechanics we have to take its unitary
representation and this gives a Gaussian propagator~\cite{4}. An
example of this problem is the Lagrangian or Hamiltonian of the
harmonic oscillator.

Thus we were led to the deuteron which we can consider, as is
usually done in nuclear physics, as a system of neutron and proton,
of essentially the same mass $m$ and interacting through an harmonic
oscillator potential of frequency $\omega$.

With units in which $\hbar = m = c=1$ ($c$ being the velocity
of light so that everything will be dimensionless) and if we apply
suddenly at $t=0$ an electrostatic field, in the direction~$z$,
acting solely on the proton, the Lagrangian will be
        \begin{gather*}
        {\mathcal L} = \frac{1}{2} \big(\dot {{\boldsymbol r}}_1+ \dot {{\boldsymbol r}}^2_2\big)
        - \frac{1}{2} \omega^2 ({\boldsymbol r}_1 - {\boldsymbol r}_2)^2 - {\mathcal E} z_1,
        \label{me04}
        \end{gather*}
where the indexes 1 and 2 will correspond respectively to the proton
and neutron and ${\mathcal E}$ is in our units the intensity of a linear
potential between the plates of a condenser multiplied by the charge
of the proton.

We proceed now to consider an orthogonal transformation leading
essentially to relative and center of mass coordinates
                \begin{gather*}
                {\boldsymbol r} = \frac{1}{\sqrt2} ({\boldsymbol r}_1 - {\boldsymbol r}_2), \qquad
                {\boldsymbol R}= \frac{1}{\sqrt2} ({\boldsymbol r}_1 + {\boldsymbol r}_2)
        \label{me05}
        \end{gather*}
which transforms ${\mathcal L}$ to
        \begin{gather}
          {\mathcal L} = \left( \frac{1}{2} \dot {{\boldsymbol r}}^2 - \frac{1}{2} \omega^2
          {\boldsymbol r}^2 - \frac{{\mathcal E}}{\sqrt2} z\right) + \left(\frac{1}{2} \dot{{\boldsymbol R}}^2 - \frac{{\mathcal E}}{\sqrt2} Z\right),
        \label{me06}
        \end{gather}
where the components of the vectors ${\boldsymbol r}$ and ${\boldsymbol R}$ are indicated
respectively as
     \begin{gather*}
      {\boldsymbol r} = {\boldsymbol i} x + {\boldsymbol j} y + {\boldsymbol k} z, \qquad
      {\boldsymbol R} = {\boldsymbol i} X + {\boldsymbol j} Y + {\boldsymbol k}  Z
        \label{me07}
        \end{gather*}
with ${\boldsymbol i}$, ${\boldsymbol j}$, ${\boldsymbol k}$ being unit vectors in the directions
indicated.

If a Lagrangian can be expressed as a sum of two terms depending on
different observables the propagator is given by the product of the
propagators associated with these two terms. Furthermore, as the
squares of the vectors ${\boldsymbol r}$, $\dot {{\boldsymbol R}}$ are given by
         \begin{gather*}
        r^2 = x^2 + y^2 + z^2, \qquad \dot R^2 = \dot X^2 + \dot Y^2
        + \dot Z^2
 \label{me08}
             \end{gather*}
it is convenient to express our problem in cartesian coordinates
where the Lagrangian \eqref{me06} becomes
         \begin{gather}
    {\mathcal L} = \left[\frac{1}{2} \big(\dot x^2 - \omega^2 x^2\big) + \frac{1}{2} \big(\dot
    y^2 - \omega^2 y^2\big) + \frac{1}{2} \big(\dot z^2 - \omega^2 z^2 - \sqrt2
    {\mathcal E} z\big) \right]\nonumber\\
\phantom{{\mathcal L} = }{}    + \left[ \frac{1}{2} \dot X^2 + \frac{1}{2} \dot Y^2 + \frac{1}{2} \big(
    \dot Z^2 -\sqrt2 {\mathcal E} Z\big)\right].
 \label{me09}
         \end{gather}

 Thus, from the observation at the beginning of the previous
 paragraph, the propagator associated with ${\mathcal L}$ will be the
 product of the one dimensional propagators associated with the
 six terms appearing in \eqref{me09}. For $\frac{1}{2} (\dot x^2 -
 \omega^2x^2)$ the propagator is given in \cite[p. 178, formula (6.2.33)]{2}  as
                         \begin{gather}
        \left[ \frac{\omega}{2\pi i \sin \omega t} \right]^\frac{1}{2} \exp \left\{
        -\frac{\omega}{2i} \left[ \big(x^2 + x'^2\big) \cot \omega t -
        \frac{2xx'}{\sin \omega t} \right]\right\}.
 \label{me10}
 \end{gather}

For $\frac{1}{2} (\dot y^2- \omega^2 y^2)$ the propagator is again of the
form \eqref{me10} but $x$, $x'$ replaced by $y$, $y'$. For $\frac{1}{2} (\dot
z^2-\omega^2 z^2-\sqrt2 {\mathcal E} z)$ we complete the square introducing
the  variable $\bar z$ by
\begin{gather}
            \bar z = z + \big({\mathcal E} /\sqrt2 \omega^2\big)
 \label{me11}
 \end{gather}
 and thus the one dimensional Lagrangian becomes
 \begin{gather}
    \frac{1}{2} \big(\dot{\bar z}^2 - \omega^2 \bar z^2\big) +
    \frac{{\mathcal E}^2}{4\omega^2}
 \label{me12}
 \end{gather}
 as $\dot{\bar z}=\dot z$. Furthermore for the constant term
 $({\mathcal E}^2/4\omega^2)$ we can apply the relation between
 propagators and Green functions of the time dependent
 Schr\"odinger equation, mentioned in p.~2 of reference~\cite{2}, to
 see that it contributes the phase term $\exp [-i
 ({\mathcal E}^2/4\omega^2)t]$ while the remainder in~\eqref{me12} is just the
 one-dimensional Lagrangian of the oscillator for the propagator
 of which we can use equation~\eqref{me10} replacing $x$, $x'$ by $\bar z$, $\bar
 z'$. It is convenient to express the propagator involving the
 relative coordinates in terms of only barred variables if we add
 the definitions
\begin{gather*}
        \bar x = x,\qquad \bar y = y
 \label{me13}
\end{gather*}
 as in these coordinates we have only the oscillator potential for
 $x$ and $y$. With the definitions
                  \begin{gather*}
        \bar r^2 = \bar x^2 + \bar y^2 + \bar z^2, \qquad \bar{{\boldsymbol r}} =
        {\boldsymbol i} \bar x + {\boldsymbol j} \bar y + {\boldsymbol k} \bar z
 \label{14}
 \end{gather*}
and the previous discussion concerning the phase factor associated
with $({\mathcal E}^2/4\omega^2)$ we obtain then that the part of the
propagator related to the relative coordinate $\bar {{\boldsymbol r}}$ becomes
\begin{gather}
    \exp \left[ - i \big({\mathcal E}^2 /4\omega^2\big) t\right]
    \left[\frac{\omega}{2\pi i \sin \omega t}\right]^{3/2} \exp
    \left\{ -\frac{\omega}{2i} \left[ \big(\bar r^2 + \bar r'^2\big) \cot
    \omega t- \frac{2\bar{{\boldsymbol r}} \cdot \bar{{\boldsymbol r}}'}{\sin \omega
    t}\right]\right\}.
 \label{me15}
         \end{gather}

Turning now to the expression \eqref{me09} with capital letters we
start with $\frac{1}{2} \dot X^2$ which from refe\-ren\-ce~\cite[p.~174,
formula (6.2.10)]{2}  becomes
\begin{gather}
    \frac{1}{(2\pi i t)^\frac{1}{2}} \exp \left[ \frac{i}{2t}
    (X-X')^2\right].
         \label{me16}
     \end{gather}

 For $\frac{1}{2} \dot Y^2$ it is the same formula \eqref{me16} but
 with $X$, $X'$ replaced by $Y$, $Y'$. For $(\frac{1}{2} \dot
 Z^2-\frac{{\mathcal E}}{\sqrt2} Z)$ we use reference~\cite[p.~175,
 formula (6.2.18)]{2}  to get
        \begin{gather}
   \frac{1}{(2\pi i t)^{\frac{1}{2}}}  \exp \left\{ i
   \left[\frac{(Z-Z')^2}{2t} - \frac{{\mathcal E} t}{2\sqrt2} (Z+Z') -
   \frac{{\mathcal E}^2t^3}{48}\right]\right\}.
 \label{me17}
 \end{gather}

 Multiplying \eqref{me16}, the corresponding expression for $Y$,
 $Y'$, and \eqref{me17} we get for the center of mass part of the
 Lagrangian the expression
\begin{gather}
    \frac{1}{(2\pi i t)^{3/2}}  \exp \left\{ i
   \left[\frac{({\boldsymbol R}- {\boldsymbol R}')^2}{2t} - \frac{{\mathcal E} t}{2\sqrt2} (Z+Z') -
   \frac{{\mathcal E}^2t^3}{48}\right]\right\}.
 \label{me18}
         \end{gather}

 Now the full propagator associated with the Lagrangian
 \eqref{me09} will be the product of \eqref{me15} and~\eqref{me18}.

 For a deuteron at $t>0$ the ground state is
        \begin{gather}
        A \exp \left(-\frac{1}{2} \omega r^2\right) \exp (i {\boldsymbol K} \cdot {\boldsymbol R}),
 \label{me19}
 \end{gather}
where $A$ is the normalization of the Gaussian term given by
        \begin{gather*}
    A = (\omega/\pi)^{3/4}.
 \label{me20}
 \end{gather*}

If we want to know these wave function at time $t$ we have to apply
the propagator associated to \eqref{me09} to the initial state
\eqref{me19} which we proceed to do in the next section.

\section[The wave function for the deuteron at $t>0$ in a
suddenly applied electrostatic field]{The wave function for the deuteron at ${\boldsymbol t>0}$\\ in a
suddenly applied electrostatic field}


As we have obtained now the propagator \eqref{me15}, \eqref{me18} associated
with the Lagrangian \eqref{me09}, we can apply it to the state
\eqref{me19} to get through equation \eqref{me02} the state at a
given time $t>0$.

The calculation involves integrals of exponentials of quadratic
expressions in the variables~$\bar {{\boldsymbol r}}'$ and ${\boldsymbol R}'$ which can be
carried out by completing squares in the exponents. We will
calculate explicitly one example and then give the final result for
$\psi ({\boldsymbol r}, {\boldsymbol R}, t)$ when $t>0$.

Remembering that for the $x$ component of the relative vector ${\boldsymbol r}$
we have the propagator \eqref{me10}, when we apply it to the part of
the initial wave function $\exp(-\frac{1}{2} \omega x^2)$ of \eqref{me19}
we have to evaluate the integral
                \begin{gather}
                \int^\infty_{-\infty} A^{1/3}
                \left(\frac{\omega}{2\pi i\sin\omega
                t}\right)^{1/2} \exp \left\{ \frac{i \omega}{2} \left[
                \big(x^2 + x'^2\big) \cot \omega t - 2 xx'(\sin \omega
                t)^{-1}\right]\right\} e^{-\frac{1}{2} \omega x'^2}
                dx'\nonumber\\
                \qquad{}=A^{1/3} \left( \frac{\omega}{2\pi i\sin \omega
                t}\right)^\frac{1}{2} \exp \left(\frac{i\omega}{2} x^2 \cot \omega
                t\right) \int^\infty_{-\infty} e^{-\beta x'^2 + \alpha
                x'} dx',
 \label{me21}
         \end{gather}
where
        \begin{gather}
    \beta= \frac{\omega}{2} (1-i \cot \omega t), \qquad \alpha = - i x\omega
    (\sin \omega t)^{-1}
 \label{me22}
 \end{gather}
and the real part of $\beta$ is positive. From reference~\cite{5} we
have that
         \begin{gather}
    \int^\infty_{-\infty} \exp \big(-\beta x'^2 + \alpha x'\big) dx' =
    (\pi/\beta)^\frac{1}{2} \exp \big(\alpha^2/4\beta\big)
 \label{me23}
 \end{gather}
and from \eqref{me22}  we obtain
         \begin{gather*}
    \beta= \frac{\omega}{2i} (\sin \omega t)^{-1} \exp (i \omega t),\qquad
    \frac{\alpha^2}{4\beta} = i \frac{\omega}{2} x^2 \cot \omega t -
    \frac{1}{2} \omega x^2
 \label{me24}
 \end{gather*}
so that replacing in  \eqref{me23} and then in \eqref{me21} we
obtain
         \begin{gather}
    A^{1/3} \exp \left(-\frac{1}{2} \omega x^2\right) \exp \left(-\frac{i \omega t}{2}\right)
     \label{me25}
 \end{gather}
which, as we should expect is, in our units, the phase term
associated with the energy $(\omega/2)$ of the ground state of the
oscillator.

The same result holds for the $y$ variable and both $x$, $y$ can be
replaced by $\bar x$, $\bar y $, as the electrostatic potential is
applied only in the $z$ direction.

For the case when our initial wave function is $A^{1/3} \exp
(-\omega z^2/2)$, it is best to replace in it $z$ by $[\bar z-
({\mathcal E} /\sqrt2 \omega^2)]$, as indicated in \eqref{me11}, and use
for the variable $\bar z$ the propagator of the oscillator to which
only a time dependent phase factor is multiplied as discussed after
equation  \eqref{me12}.

With transition to the center of mass coordinates of the deuteron,
denoted by capital letters, the propagator is just that of the free
particle when we have $X$, $Y$ so applying it to the corresponding
part of the plane wave $\exp (i K_X)$, $\exp (i K_y Y)$ they will only
give the phase factor
\begin{gather*}
    \exp (i K_x X) \exp \big(-i K^2_x t/2\big), \qquad \exp (iK_y Y) \exp \big(-i
    K^2_y t/2\big).
 \label{me26}
 \end{gather*}

For the $Z$ center of mass coordinate we have to apply to $\exp (i
K_z Z)$ the propagator  \eqref{me17} and the evaluation of the
integral again follows procedures similar to those indicated in
equations~\eqref{me21}  to \eqref{me25}.

Combining then all our results we can say that for $t>0$ our wave
function $\psi ({\boldsymbol r},{\boldsymbol R}, t)$ will become
   \begin{gather}
               \psi ({\boldsymbol r}, {\boldsymbol R}, t) = A \left(e^{-\frac{1}{2} m \omega
              r^2} e^{-i3/2 \omega t}\right) \exp \left[ i \left({\boldsymbol K}\cdot
              {\boldsymbol R} - \frac{1}{2} K^2 t\right)\right] \exp \left(-\frac{i {\mathcal E}^2 t}{4\omega^2}\right)\nonumber\\
\phantom{\psi ({\boldsymbol r}, {\boldsymbol R}, t) =}{}\times
              \exp \left\{
              \frac{{\mathcal E}}{\sqrt2 \omega} (e^{-i\omega t} - 1) z +
              \frac{{\mathcal E}^2}{2\omega^3} \left[e^{-i\omega t} -\frac{1}{2}
              (1+\cos^2\omega t)\right] + \frac{i\sin2\omega
              t}{4}\right\}\nonumber\\
\phantom{\psi ({\boldsymbol r}, {\boldsymbol R}, t) =}{}\times
              \exp \left\{ - i \left[ \frac{{\mathcal E} t^3}{12} + K_z
              \frac{{\mathcal E} t^2}{2\sqrt2} - \frac{Z{\mathcal E}
              t}{\sqrt2}\right]\right\}.
 \label{me27}
 \end{gather}

\section[The probability density for the deuteron  in a
suddenly applied electrostatic field]{The probability density for the deuteron\\  in a
suddenly applied electrostatic field}

If we turn to the ``Diffraction in time'' problem  discussed in the
Appendix, we note that to get information on its time dependent
behavior we need not the wave function but its absolute value
squared i.e. the probability density.  Once we have this we can
discuss the behavior in time with the help of the Cornu
spiral~\cite{3}. This of course is a general procedure in quantum
mechanics and thus starting from the wave function in \eqref{me27}
we need to write down $|\psi ({\boldsymbol r}, {\boldsymbol R}, t)|^2$.

We see immediately that all part associated with the center of
mass coordinates, i.e. capital $ X$,  $Y$, $Z$, disappears, as it would
also disappear for the wave function at time $t=0$ given by~\eqref{me19}.

For the relative coordinates we have to keep in the exponent of
\eqref{me27} only the terms that are real multiplied by 2, so we get
           \begin{gather*}
        |\psi ({\boldsymbol r}, {\boldsymbol R}, t)|^2 = A^2 \exp \left\{ - \omega r^2
        - \frac{2{\mathcal E}}{\sqrt2 \omega} (1-\cos \omega t)              -
        \frac{{\mathcal E}^2}{2\omega^3} \big[ -2\cos \omega t +
              \big(1+\cos^2 \omega t\big)\big]\right\}\nonumber\\
\phantom{|\psi ({\boldsymbol r}, {\boldsymbol R}, t)|^2 }{}
              =A^2 \exp \big[\!-\omega \big(x^2\! + y^2\big)\big] \!\exp \left\{
              -\omega \!\left[ z^2\! + \! \frac{2{\mathcal E}}{\sqrt2 \omega}(1-\cos
              \omega t) z + \!\frac{{\mathcal E}^2}{2\omega^4} (1-\cos \omega
              t)^2\right]\!\right\}\!\!\! \nonumber\\
\phantom{|\psi ({\boldsymbol r}, {\boldsymbol R}, t)|^2 }{}
  =A^2 \exp \big[-\omega \big(x^2 + y^2\big)\big] \exp \left\{
              -\omega \left[ z + \frac{{\mathcal E}}{\sqrt2 \omega}(1-\cos
              \omega t) \right]^2\right\} \nonumber\\
\phantom{|\psi ({\boldsymbol r}, {\boldsymbol R}, t)|^2 }{}
    =A^2 \exp \big[-\omega \big(x^2 + y^2\big)\big] \exp \left\{
              -\omega \left[ z+ \frac{\sqrt2 {\mathcal E}}{\omega}\sin^2 (
              \omega t/2)\right]^2\right\},
 \label{me28}
 \end{gather*}
where used a trigonometric relation for the last expression.

We have thus a very simple transient behavior that depends only on
the relative coordinates and in which $x$, $y$ have the standard
Gaussian behavior of range $\omega^{-1/2}$ in our units, while $z$
oscillates around this range with an amplitude $(\sqrt2
{\mathcal E}/\omega^2) \sin^2 (\omega t/2)$. The period of oscillation is
         \begin{gather*}
    (\omega T/2) = \pi \qquad \mbox{or}\qquad  T=(2\pi/\omega)
 \label{me29}
     \end{gather*}
while the amplitude  is $(\sqrt2 {\mathcal E}/\omega^2)$.

Note that coordinates of proton and neutron, in the direction of the
electrostatic field, are given respectively by
        \begin{gather*}
    z_1 = \frac{1}{\sqrt2} (Z + z), \qquad z_2 = \frac{1}{\sqrt2} (Z-z)
 \label{me30}
 \end{gather*}
so if we take our origin at the position of the center of mass, the
distance between proton and neutron goes as $\sqrt2 z$ and thus the
deuteron is vibrating with an amplitude proportional to $(2 {\mathcal E}/
\omega^2) \sin^2 (\omega t/2)$ and it could radiate for strong
electrostatic fields.

Clearly for very strong electrostatic fields the deuteron could
desintegrate and this information can be obtained also from the
discussion of the classical version of the problem.

The information available for propagators in reference~\cite{2} would
allow us to discuss also the problem of a time dependent
electrostatic field which reflect more the physical situation. The
formulas we could use (6.2.42) is in p.~180 of reference~\cite{2}.

For a constant electromagnetic field given by a vector potential ${\boldsymbol A}$
we could use formula (6.2.17) of p.~175 of reference~\cite{2}.

In general Feynman procedure gives a powerful tool for analyzing
transient phenomena when external fields are applied to bound
systems.

\section{General procedure for dealing with transient phenomena\\
caused by a sudden perturbation}

While the propagator $K({\boldsymbol x}, t; {\boldsymbol x}')$ is very useful for the
discussion of transient phenomena, it is not usually available for
the general potential $V_+ ({\boldsymbol x})$. It is thus useful to develop
an approximate procedure that is always available.

Denoting for $H_\pm$ the Hamiltonians associated with the potential
$V_\pm$ we have that
              \begin{gather*}
        H_- = T + V_-, \qquad H_+ = T + V_+ =  H_- + (V_+ - V_-)
        \label{me31}
        \end{gather*}
where $T$ is the kinetic energy.

We shall assume that $H_-$ has only a discrete spectrum. In the case it
also has a continuous spectrum we can reduce it to the previous case
by inserting the whole system into a box. We shall denote the energies
and eigenfunctions of $H_-$ as
\begin{gather*}
            H_- \phi_n = E_n \phi_n,
        \label{me32}
        \end{gather*}
where
\begin{gather*}
            E_0 \leq E_1 \leq E_2 \leq \cdots \leq E_{n-1} \leq
            E_{n} \leq E_{n+1}\cdots.
        \label{me33}
        \end{gather*}
For $H_+$ we are in search of a solution $\psi({\boldsymbol x}, t)$ of the time
dependent equation (in c.g.s. units)
              \begin{gather}
                i\hbar \frac{\partial \psi}{\partial t} = H_+ \psi =
                [H_- + (V_+ - V_-)] \psi
        \label{me34}
        \end{gather}
with the initial value
            \begin{gather*}
        \psi ({\boldsymbol x}, 0) = \phi_\nu ({\boldsymbol x}),
        \label{me35}
        \end{gather*}
where $\nu$ is some integer.

We define the Laplace transform of $\psi ({\boldsymbol x}, t)$ by
              \begin{gather*}
        \bar \psi ({\boldsymbol x}, s) = \int^\infty_0 e^{-st} \psi ({\boldsymbol x}, t) dt.
        \label{me36}
        \end{gather*}
Applying the transform to equation \eqref{me34} we get
        \begin{gather}
          i \hbar \int^\infty_0 e^{-st} \frac{\partial \psi}{\partial t}dt
          = i\hbar \int^\infty_0 \frac{\partial}{\partial t}(e^{-st} \psi({\boldsymbol x}, t)) dt + i \hbar s
          \bar \psi ({\boldsymbol x}, s)\nonumber\\
          \phantom{i \hbar \int^\infty_0 e^{-st} \frac{\partial \psi}{\partial t}dt}{}
          = - i \hbar \psi ({\boldsymbol x}, 0) + i\hbar s \bar\psi ({\boldsymbol x},s) = [H_-
          + (V_+ - V_-)] \bar \psi ({\boldsymbol x},s).
                  \label{me37}
        \end{gather}

Now the eigenstates $\phi_n$ of $H_-$ are a complete orthonormal set
so we can make the expansion
        \begin{gather*}
               \bar \psi ({\boldsymbol x}, s) = \sum^\infty_{n=0} a_n (s) \phi_n ({\boldsymbol x})
        \label{me38}
        \end{gather*}
so that equation \eqref{me37} becomes
     \begin{gather*}
            -i\hbar \phi_\nu ({\boldsymbol x}) + i\hbar s \sum_n a_n (s) \phi_n
            ({\boldsymbol x})
            = [ H_- + (V_+ - V_-)] \sum_n a_n (s) \phi_n ({\boldsymbol x}).
        \label{me39}
        \end{gather*}

Multiplying by $\psi^*_{n'} ({\boldsymbol x})$ and integrating over the
variables ${\boldsymbol x}$ we get
\begin{gather}
               -i\hbar \delta_{n'\nu} + i \hbar s a_{n'} (s)
                              = E_{n'} a_{n'} (s) + \sum_n \langle n'|V_+ - V_-| n\rangle a_n (s),
        \label{me40}
        \end{gather}
where
                    \begin{gather*}
            \langle n' |V_+ - V_-| n\rangle  = \int \phi^*_{n'}({\boldsymbol x}) [ V_+ ({\boldsymbol x}) -
            V_- ({\boldsymbol x}) ] \phi_n ({\boldsymbol x}) d{\boldsymbol x}.
        \label{me41}
        \end{gather*}

The equation \eqref{me40} corresponds to an infinite set of linear
algebraic equations for the coefficients~$a_n (s)$. We can solve
them with the usual assumption that after a given index $N$ all
coefficients vanish, and then we get the $a_n(s)$ for $n=0,1,\dots, N$
from which we can write the $\bar \psi ({\boldsymbol x}, s)$ as
\begin{gather*}
            \bar \psi ({\boldsymbol x}, s) = \sum^N_{n=0} a_n (s) \phi_n ({\boldsymbol x}).
        \label{me42}
        \end{gather*}

We want though to determine $\psi({\boldsymbol x},t)$ for which we need the
reciprocal of the Laplace transform
\begin{gather*}
                \psi ({\boldsymbol x},t) = \frac{1}{2\pi i}
                \int^{c+i\infty}_{c-i\infty} \bar \psi ({\boldsymbol x},s) e^{st}
                ds
                = \frac{1}{2\pi \hbar} \int^{\infty+i \hbar
                c}_{-\infty + i \hbar c} \exp (-i Et/\hbar)
                \bar \psi ({\boldsymbol x}, -i E/\hbar) dE,
        \label{me43}
        \end{gather*}
where we replace the variable $s$ by
                                \begin{gather*}
                s= -i (E/\hbar)
        \label{me44}
        \end{gather*}
with $E$ being integrated along a line above all poles of the
function $\bar \psi ({\boldsymbol x}, -i E \hbar^{-1})$. Note that the contour
can be completed with a circle from below and get essentially the
residues at values of $E$ given by the homogeneous set of linear
equations
\begin{gather*}
                (E-E_{n'}) a_{n'} = \sum_n \langle n'| V_+ - V_-| n\rangle a_n
        \label{me45}
        \end{gather*}
or \begin{gather*}
                E a_{n'} = \sum_n \langle n'| H_+ | n\rangle a_n
        \label{me46}
        \end{gather*}
which essentially give us the energies associated with the levels of
the Hamiltonian $H_+$.

\appendix


\section{Diffraction in time}

Units
\begin{gather*}
\hbar = m=c=1,\\
 u (l) = (\hbar/mc), \qquad u (t) = (\hbar/mc^2),\qquad
  u (m) = m.
  \end{gather*}
Free particle propagator
\begin{gather*}
K (x,t; x', 0) = (2\pi i t )^{-\frac{1}{2}} \exp \big[ (i/2t) (x-x')^2\big].
\end{gather*}
The integral for diffraction in time is
\begin{gather*} M (x,k,t) = \int^0_{-\infty} K (x,t; x', 0) \exp (ikx') dx',\\
\phantom{M (x,k,t)}{}= (2\pi it)^{-\frac{1}{2}} \exp \left\{ i \left(kx-\frac{1}{2} k^2 t\right)\right\}
\int^0_{-\infty} \exp \left\{ \frac{i}{2t} [ x'+(kt-x)]^2\right\} dx',
\end{gather*}
where the last expression comes from completing the square for $x'$.

We introduce the variable $u$ by the definition
\begin{gather*}
(\pi/2)^\frac{1}{2} u = (2t)^{-\frac{1}{2}} [x' + (kt-x)]
\end{gather*}
from which it follows that for $x'=0$, we have $u=u_0$
\begin{gather*}
 u_0= (\pi t)^{-\frac{1}{2}} (kt-x), \qquad \sqrt{\pi/2} du = (2t)^{-\frac{1}{2}}
 dx
 \end{gather*}
and thus $M(x,k,t)$  becomes
\begin{gather*} M (x,k,t)= e^{-i\pi/4} \exp\left[i
\left(kx-\frac{1}{2} k^2 t\right)\right] \left[ \int^0_{-\infty} \exp (i \pi u^2/2) du +
\int^{u_0}_0 \exp (i\pi u^2/2) du \right].
\end{gather*}

But we have~\cite{5}
\begin{gather*}
    \int^0_{-\infty}\exp (i \pi u^2/2) du = \frac{(1+i)}{2}, \qquad
    \int^{u_0}_0 \exp (i \pi u^2/2)du = C (u_0)+ i S (u_0)
    \end{gather*}
 with the definition of Fresnel integrals
 \begin{gather*} C (u_0)=
\int^{u_0}_0 \cos (\pi u^2/2) du, \qquad S (u_0)= \int^{u_0}_0 \sin
(\pi u^2/2) du,
\\
               M (x,k,t) = e^{-i\pi/4} \exp \left[i\left(kx - \frac{1}{2} k^2 t\right)\right]
\left\{ \left[ \frac{1}{2} + C(u_0)\right] + i \left[\frac{1}{2} + S
(u_0)\right]\right\}
\end{gather*}
and
\begin{gather}
|M (x,k,t)|^2 = \left\{  \left[\frac{1}{2} + C(u_0)\right]^2 +
\left[\frac{1}{2} + S (u_0)\right]^2\right\}
 \label{me58}
 \end{gather}
for which the Cornu spiral can be used to show the difference
between the quantum behavior given by \eqref{me58} and the classical
one in which the probability density is 1 in the interval
$\infty\leq x\leq vt$ and 0 for $v t \leq x\leq \infty$~\cite{3}.

\LastPageEnding

\end{document}